\def\nk{n_{\rm b}}

\def\rfr#1{eq. (\ref{#1})}

\def\virg#1{``#1''}

\def\eqi{\begin{equation}}
\def\eqf{\end{equation}}
\def\eqia{\begin{eqnarray}}
\def\eqfa{\end{eqnarray}}

\def\rp#1#2{{#1\over#2}}
\def\lb#1{\label{#1}}

\def\ton#1{\left(#1\right)}

\documentclass[Galaxies,letter,accept,oneauthor,dvipdfm,12pt,a4paper]{mdpi} 

\lastpage{x}
\doinum{10.3390/------}
\pubvolume{xx}
\pubyear{2013}
\history{Received: xx / Accepted: xx / Published: xx}
\usepackage{amsmath,amssymb,amsthm,amscd,latexsym}
\usepackage{hyperref}
\usepackage[toc,title,titletoc]{appendix}
\usepackage{graphicx,epsfig}
\RequirePackage{color}

\Title{The lingering anomalous secular increase of the eccentricity of the orbit
of the Moon: further attempts of explanation of cosmological origin}

\Author{Lorenzo Iorio $^{1,}$*}

\address{%
$^{1}$ Italian Ministry of Education, University and Research (M.I.U.R.)-Education, Fellow of the Royal Astronomical Society (F.R.A.S.), Viale Unit\`{a} di Italia 68, 70125, Bari (BA), Italy. Tel. +39 3292399167}

\corres{lorenzo.iorio@libero.it}

\abstract{
A new analysis of extended data records collected with the Lunar Laser Ranging (LLR) technique performed with improved tidal models was not able to resolve the issue of the anomalous rate $\dot e$ of the eccentricity $e$ of the orbit of the Moon, which is still in place with a magnitude of $\dot e=(5\pm 2)\times 10^{-12}$ yr$^{-1}$.
Some possible cosmological explanations are offered in terms of the post-Newtonian effects of the cosmological expansion, and of the slow temporal variation of the relative acceleration rate $\ddot{S} S^{-1}$ of the cosmic scale factor $S$. None of them is successful since their predicted secular rates of the lunar eccentricity are too small by several orders of magnitude.
}



\begin{document}

\section{Introduction}\lb{Introduzione}

In 2009, Williams and Boggs \citep{WilliamsBoggs2009} reported  an anomalous secular rate of the eccentricity $e$ of the orbit of the Moon
\eqi\dot e = (9\pm 3)\times 10^{-12}\ {\rm yr^{-1}}.\lb{rate} \eqf They fit the dynamical models of the DE421 ephemerides \citep{DE421} to a record of ranges collected with the Lunar Laser Ranging (LLR) technique from March 1970 to November 2008. The effect of \rfr{rate} is anomalous in the sense that it is in excess with respect to the eccentricity rate predicted with the tidal models of the DE421 ephemerides.
Anderson and Nieto \citep{2010IAUS..261..189A} included \rfr{rate} in the current astrometric anomalies in the Solar System.
Some more or less sound attempts to find an explanation of \rfr{rate} in terms of either standard and non-standard gravitational physics  have been performed so far
\citep{2011MNRAS.415.1266I, 2011AJ....142...68I, 2013PhyEs..26...82Z, 2013AdSpR..52.1297A, 2013PhyEs..26..567A}.

Recently, Williams et al. \citep{WilliamsetalPS2014} extended their analysis of the LLR data from March 1970 to April 2013  by using the new DE430 ephemerides \citep{2014IPNPR.196C...1F} with improved tidal models. As a result, the anomalous eccentricity rate of the lunar orbit, although reduced with respect to \rfr{rate}, did not disappear, amounting now to
\eqi \dot e = (5\pm 2)\times 10^{-12}\ {\rm yr^{-1}}\lb{rate2}.\eqf
The rate of \rfr{rate2} exhibited a low correlation with other solved-for estimated parameters \citep{WilliamsetalPS2014}.

In this Letter, we propose to look for further possible physical mechanisms able to explain the lingering anomalous lunar eccentricity rate.
\section{Ruling out some possible mechanisms of cosmological origin}\lb{cosmo}

At the Newtonian level, the cosmological expansion induces a Hooke-type two-body acceleration proportional to the binary's separation $r$ through an \virg{elastic} constant $\mathcal{K}$ given by the relative  acceleration rate $\ddot{S}S^{-1}$ of the cosmic scale factor $S$; see, e.g., \citep{2010RvMP...82..169C} and references therein. Such an acceleration, proportional to the square of the Hubble parameter $H$, affects neither the shape nor the size of the orbit of a localized binary system, as it was calculated by several authors with a variety of different approaches \citep{1998ApJ...503...61C, 2007CQGra..24.5031M, 2007PhRvD..75f4031S, 2007PhRvD..75f4011A, 2012MNRAS.422.2945N}.

At the first post-Newtonian (pN) level, a velocity-dependent acceleration linear in $H$ occurs \citep{2012PhRvD..86f4004K}: in principle, it does secularly change both the semimajor axis $a$ and the eccentricity $e$ of the orbit of a test particle moving about a central body of mass $M$ \citep{2013MNRAS.429..915I}.
The resulting eccentricity rate, averaged over one orbital period $P_{\rm b}$ of the test particle, is \citep{2013MNRAS.429..915I}
\eqi\dot e^{(H{\rm pN})} = \rp{4H_0 a^2 \nk^2\sqrt{1-e^2}\ton{-1+\sqrt{1-e^2}}}{c^2 e},\lb{dedtKop}\eqf
where $c$ is the speed of light in vacuum, $H_0$ is the value of the Hubble parameter at the present epoch, and $\nk = \sqrt{GM a^{-3}} = 2\pi P^{-1}_{\rm b}$ is the Keplerian mean motion; $G$ is the Newtonian constant of gravitation.
In the case of the Earth-Moon system, by using the latest determination for the Hubble parameter from the Planck mission \citep{2013arXiv1303.5076P}
$H_0 = 6.88\times 10^{-11}\ {\rm yr^{-1}}$,
\rfr{dedtKop} yields a (negative) lunar eccentricity rate as little as \eqi\dot e^{(H{\rm pN})} = -1\times 10^{-22}\ {\rm yr}^{-1}\lb{mini}.\eqf Apart from the sign, \rfr{mini} is ten orders of magnitude smaller than \rfr{rate2}.

Recently, in \citep{2014Galax...2...13I} it has been pointed out  that a slow temporal variation of $\ddot{S} S^{-1}$ affects the local dynamics of a two-body system by secularly changing both $a$ and $e$. In the case of matter-dominated epochs with Dark Energy \citep{2014Galax...2...13I}, it turns out that, to the first order in the power expansion of $\ddot{S} S^{-1}$, the mean eccentricity rate is
\eqi\dot e^{(\Omega_{\Lambda})} = \rp{\ton{4 + e}\ton{-1 + e^2}\mathcal{K}_1}{4\nk^2},\lb{dedtIorio}\eqf
where
\eqi\mathcal{K}_1 \doteq \rp{3}{2}H_0^3\Omega_{\Lambda}^{3/2}\coth\ton{\rp{3}{2}H_0 t_0\sqrt{\Omega_{\Lambda}}}{\rm csch}^2\ton{\rp{3}{2}H_0 t_0\sqrt{\Omega_{\Lambda}}},\lb{Kappa1}\eqf
in which $t_0$ is the present epoch, while $\Omega_{\Lambda}$ is the Dark Energy density normalized to the critical energy density.
The latest values $t_0 = 13.817$ Gyr, $\Omega_{\Lambda} = 0.685$ from Planck \citep{2013arXiv1303.5076P} and \rfr{dedtIorio}-\rfr{Kappa1} yield for the Moon
\eqi\dot e^{(\Omega_{\Lambda})} = -2\times 10^{-35}\ {\rm yr^{-1}},\eqf whose magnitude is $23$ orders of magnitude smaller that the observed anomalous eccentricity rate of \rfr{rate2}.

Thus, we can rule out any potentially viable explanation of cosmological origin for \rfr{rate2}.
\section{Conclusions}\lb{conclusioni}

As a result of the latest LLR data analysis performed with improved tidal models, it turned out that the anomalous eccentricity rate of the lunar orbit is still lingering; it amounts to $\dot e = (5\pm 2)\times 10^{-12}$ yr$^{-1}$.

The LLR analysts seem convinced that, sooner or later, a better understanding of the intricate geophysical processes of tidal origin taking place in the interiors of our planet and of its satellite will be able to fully accommodate the selenian orbital anomaly. As such, they will continue to look for conventional causes for the anomalous eccentricity rate of the Moon.

Nonetheless, as a complementary approach, the search for  causes residing outside the Earth and the Moon themselves is worth of being pursued. If unsuccessful, it could also indirectly strengthen the relevance of the efforts towards an explanation in terms of standard geophysics.
In this respect, in this Letter it has been shown  that neither the cosmological expansion at the first post-Newtonian level nor the slow temporal variation of the relative acceleration rate of the cosmic scale factor are able to explain the anomalous lunar eccentricity increase because they induce long-term rates of change of the Moon's eccentricity too small by several orders of magnitude. Such a further negative result adds to the previous series of failed attempts to find sound  non-tidal explanations appeared in the literature so far.

\bibliography{Moonbib}{}
\bibliographystyle{mdpi-arXiv}

\end{document}